Anomalous magnetotransport properties of high-quality single crystals of Weyl semimetal $WTe_2$: Sign change of Hall resistivity

Rajveer Jha[1,2,*], Ryuji Higashinaka[1], Tatsuma D. Matsuda[1], Raquel A. Ribeiro[2] and Yuji Aoki[1,†]

[1]*Department of Physics, Tokyo Metropolitan University, Hachioji, Tokyo 192-0397, Japan*
[2]*CCNH, Universidade Federal do ABC (UFABC), Santo André, SP, 09210-580, Brazil*

## Abstract:

We report on a systematic study of Hall effect using high quality single crystals of type-II Weyl semimetal $WTe_2$ with the applied magnetic field $B$//c. The residual resistivity ratio of 1330 and the large magnetoresistance of $1.5\times10^6$ % in 9 T at 2 K, being in the highest class in the literature, attest to their high quality. Based on a simple two-band model, the densities ($n_e$ and $n_h$) and mobilities ($\mu_e$ and $\mu_h$) for electron and hole carriers have been uniquely determined combining both Hall- and electrical-resistivity data. The difference between $n_e$ and $n_h$ is ~1% at 2 K, indicating that the system is in an almost compensated condition. The negative Hall resistivity growing rapidly below ~20 K is due to a rapidly increasing $\mu_h/\mu_e$ approaching one. Below 3 K in a low field region, we found the Hall resistivity becomes positive, reflecting that $\mu_h/\mu_e$ finally exceeds one in this region. These anomalous behaviors of the carrier densities and mobilities might be associated with the existence of a Lifshitz transition and/or the spin texture on the Fermi surface.

E-mail for corresponding authors: *rajveerjha@gmail.com, † aoki@tmu.ac.jp

**Rajveer Jha**
*Department of Physics*
*Tokyo Metropolitan University*
*Hachioji-shi, Minami Osawa, 1-1*
*192-0397, Tokyo -JAPAN*

Introduction:

Quantum topological materials, including graphene, Dirac- and Weyl-semimetals [1-4] have been attracting great interests recently due to novel physical properties. In Weyl semimetals, Weyl fermions with different chiralities of left or right handed are expected to play an essential role in the transport properties. Large magnetoresistance, which manifests itself in some of Weyl semimetals, is considered to be caused by high carrier mobility of Weyl fermions. Layered transition-metal dichalcogenides typically sharing the formula, $TMTe_2$, where $TM$ is a transition metal (e.g., Mo or W), are candidate materials of so-called "type II" Weyl semimetal [5]. One such example is $WTe_2$, which shows extremely large non-saturating magnetoresistance suggesting the compensation of electrons and holes [6]. Fourier transform spectra of Shubnikov–de Haas oscillations show that $WTe_2$ has four Fermi pockets, i.e., two sets of concentric electron- and hole-Fermi pockets [7–10], being consistent with band structure calculations [7]. Unsaturated magnetoresistance even in 60 T suggests that electron and hole carrier densities are highly compensated [6]. Considering that $WTe_2$ is an almost compensated semimetal, the "two-band model" [11,12] is probably an appropriate simple model to analyze the transport properties. Hall resistivity $\rho_{yx}$, which provides a complementary information on the charge carriers, can be analyzed based on the two-band model. There are three reports for the measurements and analyses of $\rho_{yx}$ [13-15]. However, because the unique set of the solutions based on the two-band model is difficult to be obtained only using $\rho_{yx}$ data, the characters of the two types of carriers have not been clarified yet. In this paper, in order to overcome this difficulty, we utilize the data of both electrical resistivity and Hall resistivity simultaneously and determine the parameters in more reliable manner.

Experimental Details:

High quality single crystals of $WTe_2$ were obtained by Te-flux growth technique. Electrical resistivity $\rho$ and Hall resistivity $\rho_{yx}$ were measured by the four probe technique using a Physical Property Measurement System (PPMS: Quantum Design) down to 2 K. The measurements were carried out in applied magnetic fields $B//c$ up to 9 T with the current $I$ flowing along the $a$ crystallographic direction. The spot-welding technique was used to attach gold wires to the samples to make electrical contact.

Results and Discussion:

Figure 1(a) displays the temperature dependences of electrical resistivity $\rho$ for $I//a$ in applied fields $B$=0 and 9 T. In zero field, ρ shows a metallic behavior. The residual resistance ratio $RRR$ = $\rho_{300K}/\rho_{2K}$=1330, being in the highest class in the literature [6,13], attests to high quality of the single crystals with only few defects and impurities. At low temperatures where the zero field resistivity $\rho(0)$ becomes small, $\rho(9T)$ is largely enhanced compared to $\rho(0)$, demonstrating that small $\rho(0)$ is necessary for the appearance of extremely large magnetoresistance effect. The temperature dependence of magnetoresistance ($MR = \frac{\Delta\rho(B)}{\rho(0)} = \frac{\rho(B)-\rho(0)}{\rho(0)}$) in 9 T is shown in Fig. 1(b). At 2 K, $MR$% reaches to $1.5\times10^6$ %, reflecting the high quality of the single crystal. The field dependence of $MR$ measured at 2 K with $I//a$ in $B//c$ is shown in Fig. 1(c). The $\rho$ increases with increasing field without any saturation tendency roughly obeying the Kohler rule with almost $B^2$ dependence. Above ~4 T, clear Shubnikov-de Hass quantum oscillations appear. The Fourier transform spectra show the existence of four Fermi pockets (not shown), the sizes of which agree quantitatively with those obtained in previous studies [7-10].

Figure 2 shows the temperature dependence of Hall resistivity ($\rho_{yx}$) measured in 9 T for $I//a$. In 9 T, $\rho_{yx}$ is always negative in the measured temperature range, suggesting electron carriers dominate over hole carriers. While $\rho_{yx}$ shows a weak temperature dependence above 50 K, it starts to decrease rapidly below 20 K.

The magnetic-field dependences of $\rho_{yx}$ are shown in Fig. 3 (a) for several fixed temperatures below 40 K. At low temperatures, the $\rho_{yx}$ vs $B$ data also show clear quantum oscillations in the high field region above ~4 T, in a similar manner with the $\rho$ vs $B$ data shown in Fig. 1(c).

Note that the field dependence of $\rho_{yx}$ is strongly nonlinear, i.e., the Hall coefficient $\rho_{yx}/B$ is not a field-independent constant. The nonlinear behavior becomes more prominent with decreasing temperature below ~20 K. Considering that $WTe_2$ is an almost compensated semimetal, one simple model to explain the observed nonlinear field dependence of $\rho_{yx}$ is the conventional two-band model [11-15]. In this model, the total conductivity tensor $\boldsymbol{\sigma}$ is the sum of electron and hole conductivity and is expressed in the complex representation,

$$\boldsymbol{\sigma} = e\left[\frac{\mu_e n_e}{1+i\mu_e B} + \frac{\mu_h n_h}{1-i\mu_h B}\right] , \qquad (1)$$

where $e$(>0) is the electronic charge, $n_e$ and $n_h$ are the carrier density, $\mu_e$ and $\mu_h$ are the carrier mobility for electrons and holes, respectively. The real and imaginary part of the resistivity tensor (**ρ**=1/**σ**) are $\rho_{xx}$(≡ρ in this manuscript) and $\rho_{yx}$. These can be expressed as:

$$\rho_{xx}(B) = \frac{(n_h\mu_h+n_e\mu_e)+(n_h\mu_e+n_e\mu_h)\mu_e\mu_h B^2}{e[(n_h\mu_h+n_e\mu_e)^2+(n_h-n_e)^2(\mu_h\mu_e B)^2]} , \qquad (2)$$

$$\rho_{yx}(B) = \frac{B[(n_h\mu_h^2-n_e\mu_e^2)+(n_h-n_e)(\mu_h\mu_e B)^2]}{e[(n_h\mu_h+n_e\mu_e)^2+(n_h-n_e)^2(\mu_h\mu_e B)^2]} . \qquad (3)$$

The observed nonlinear field dependence of $\rho_{yx}$ can be reproduced well using eq. (3). A set of the four parameters $\{n_e, n_h, \mu_e, \mu_h\}$ can be obtained by fitting the model to the experimental data. Because of the cross-correlation among those parameters, however, it is difficult to determine the unique values for the parameters only using eq. (3). To overcome this difficulty, we perform fitting combining eqs. (2) and (3) so that the model satisfies both of the $\rho_{yx}$(B) and $\rho$(B) data at the same time; this technique was not utilized in the previous studies [13-15]. In order to determine the parameters with a high level of confidence, the nonlinear behaviors in $\rho_{yx}$(B) is necessary. Because of this, the fitting is less reliable in high temperatures (see Fig. 3 (a)). In Fig. 4, we show the fitting results in $T$<10 K. Note that, since the data in 1 T<$B$<9 T are used, the obtained parameters are considered to represent those in the high-field condition. It is possible that these parameters deviate to some extent in the low field region ($B$<~1 T).

The model calculations shown in Figs. 1 (b) and 2 by open circles and in Fig. 3(a) by thin lines demonstrate that the two-band model nicely reproduces the $\rho$ and $\rho_{yx}$ data simultaneously. The determined four parameters are plotted in Figs. 4(a-c) as a function of temperature. In this temperature range, the charge compensation is almost satisfied, i.e., the difference between $n_h$ and $n_e$ is less than 3% and decreases with decreasing temperature, finally approaching ~1% at 2 K. The temperature dependent behaviors of $\mu_e$ and $\mu_h$ are completely opposite. With decreasing temperature, $\mu_e$ decreases rapidly while $\mu_h$ increases rapidly, keeping the geometric-mean mobility $(\mu_e\mu_h)^{1/2}$ increasing slowly. In this temperature range, $\mu_h/\mu_e$ rapidly increases from ~$10^{-2}$ at 8 K to 1.3 at 2 K. Considering eqs. (2 and 3), these mutually contrasting behaviors lead to the rapidly growing negative Hall resistivity below ~20 K. At 2 K, $\mu_h/\mu_e$=1.3>1, indicating that holes dominate over electrons in the transport properties. This feature actually manifests itself in the $\rho_{yx}$ data as shown below.

Figure 3 (b) shows an expanded view of the $\rho_{yx}$-vs-B data below 1 T at 2 K. As clearly demonstrated here, we have found that $\rho_{yx}$ shows a sign change, i.e., $\rho_{yx}$>0 in $B$<0.8 T. This observation suggests that holes dominate over electrons in the transport properties, i.e., $(n_h\mu_h^2 - n_e\mu_e^2) > 0$ (see eq. (3)). As shown by a thin solid line in Fig. 3(b), the $\rho_{yx}$-vs-$B$ data can be reproduced well by eq. (3).

Recent ARPES measurements [16, 17] indicate that the bands of $WTe_2$ show an anomalous temperature-dependent shift. With increasing temperature, the bands corresponding to the hole pockets move down in energy and finally they sink completely under the Fermi level leading to a Lifshitz transition around 160 K. This phenomenon, possibly remaining down to the temperature range of the present study, may account for the anomalous behaviors of $\mu_h/\mu_e$ and $n_h/n_e$, both of which continue to increase with decreasing temperature as shown in Figs. 4 (a) and (c).

The obtained values of carrier mobility shown in Fig. 4 (a) are $10^4$~$10^6$ cm$^2$/Vs, which are higher than the reported values for $WTe_2$ (e.g., $\mu_e$=1.1×$10^4$ cm$^2$/Vs and $\mu_h$=0.65×$10^4$ cm$^2$/Vs at 0.3 K [13], and $\mu_e$=1.045×$10^3$ cm$^2$/Vs and $\mu_h$=1.165×$10^3$ cm$^2$/Vs at 5 K [14]) and are reaching to the reported ultrahigh value $\mu$=9.165×$10^6$ cm$^2$/Vs at 5 K in Dirac semimetal $Cd_3As_2$ (*RRR* = 4100) [3]. In $WTe_2$, due to the strong spin-orbit coupling, the spin texture is formed on the Fermi surface [18]. This feature, which may affect scattering processes of electrons and holes, is neglected in the present two-band model. This point might be necessary to be considered to understand the anomalous behaviors of the carrier densities and mobilities appearing in Fig. 4.

In summary, we have measured Hall resistivity of type-II Weyl semimetal $WTe_2$ in the applied magnetic field *B//c* with *I//a*. The residual resistivity ratio of 1330 and the large magnetoresistance of 1.5×$10^6$ % in 9 T at 2 K, being in the highest class in the literature, attest to their high quality. Based on a simple two-band model, the densities ($n_e$ and $n_h$) and mobilities ($\mu_e$ and $\mu_h$) for electron and hole carriers have been uniquely determined combining both Hall- and electrical-resistivity data. The negative Hall resistivity growing rapidly below ~20 K is due to a rapidly increasing $\mu_h/\mu_e$ approaching one in an almost compensated condition ($n_e \cong n_h$). Below 3 K in a low field region, we found the Hall resistivity becomes positive, reflecting that $\mu_h/\mu_e$ finally exceeds one in this region. The anomalous temperature dependences of the carrier motilities might be associated with the temperature-dependent hole-band energy shift (possibly associated with the putative Lifshitz transition) and/or anomalous carrier scattering process due to the spin texture formed on the Fermi pockets.

Acknowledgments:
We gratefully appreciate Prof. Hideyuki Sato for fruitful discussions. This work was financially supported by JSPS KAKENHI Grant Numbers JP15H03693, JP15H05884, JP16F16028 and JP16K05454.

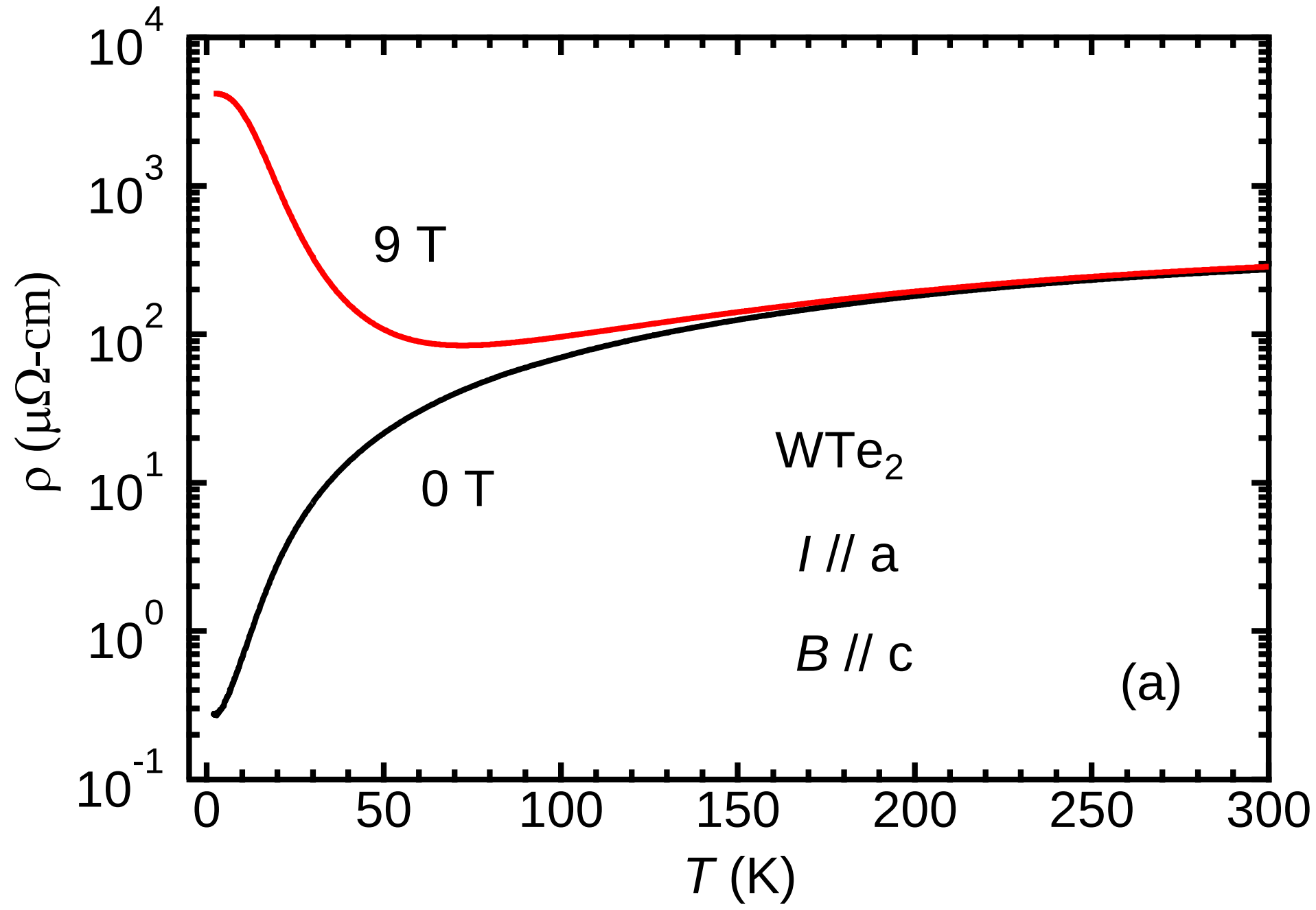

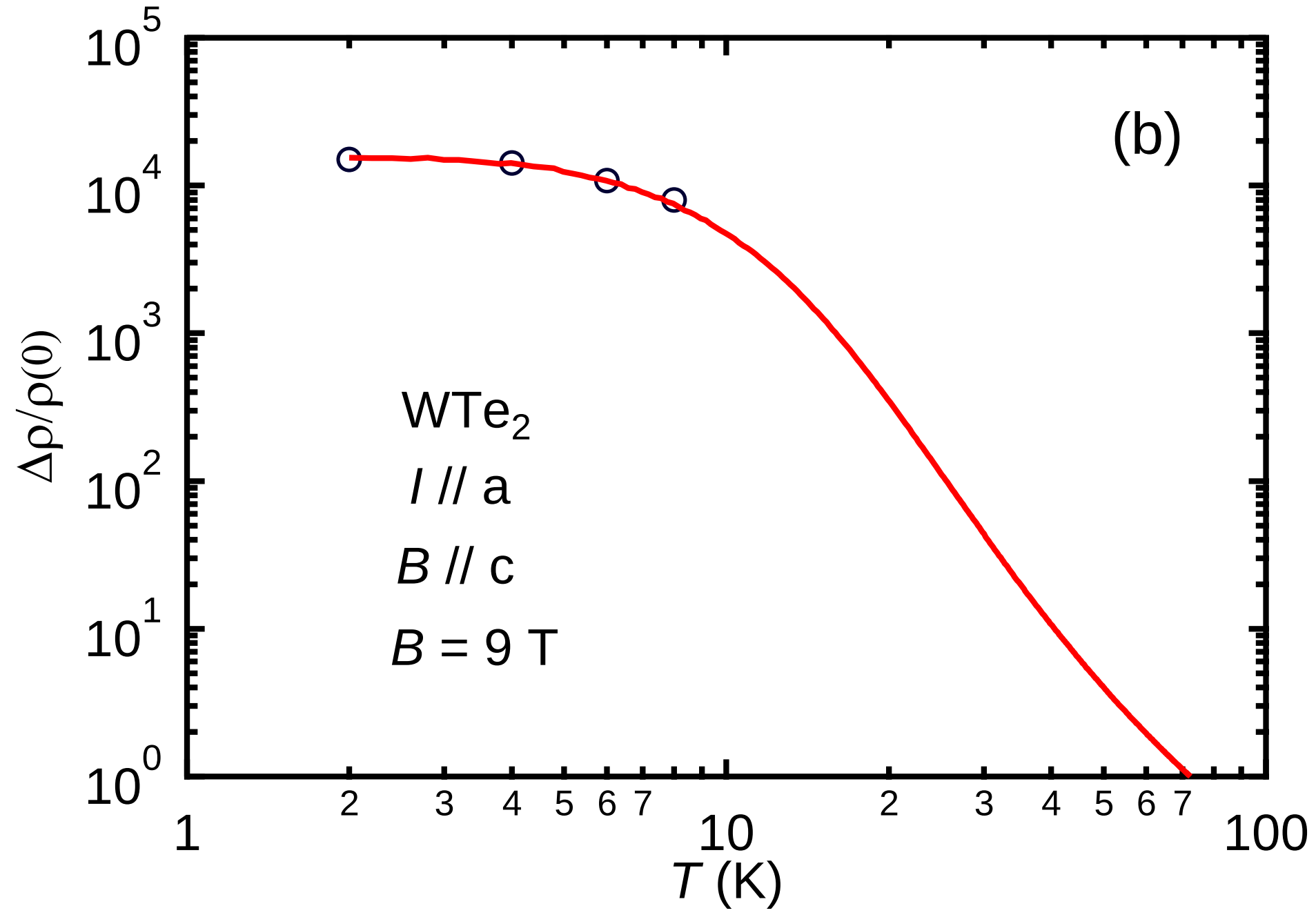

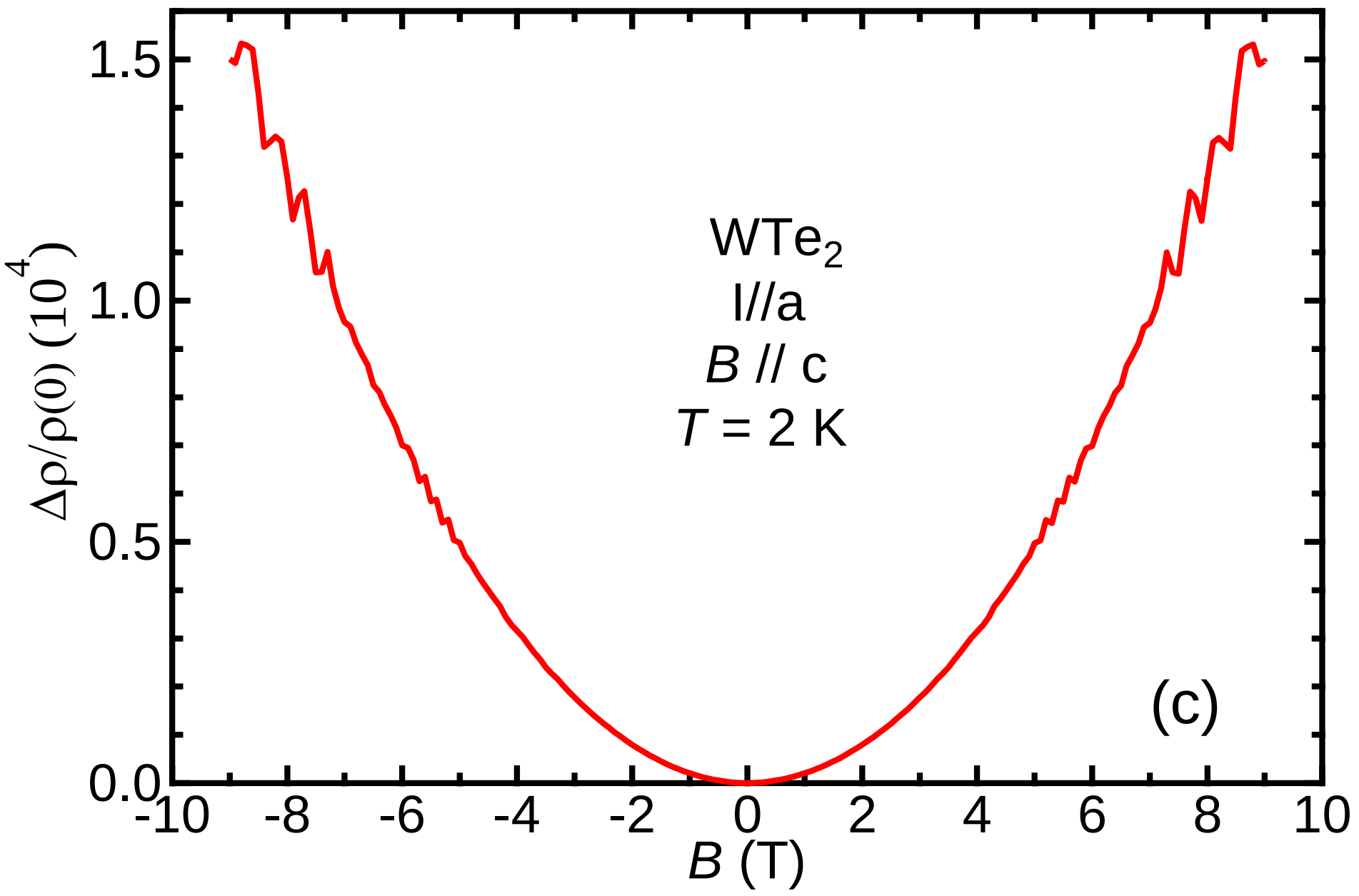

**Figure 1:** (a) Temperature dependence of the electrical resistivity $\rho$ of $WTe_2$ single crystal for the current *I*//*a* in applied fields *B*//c (0 and 9 T). (b) Temperature dependence of *MR*. Two-band model fitting represented by open circles shows a good agreement with the MR data. (c) Field dependent *MR* at *T*=2 K.

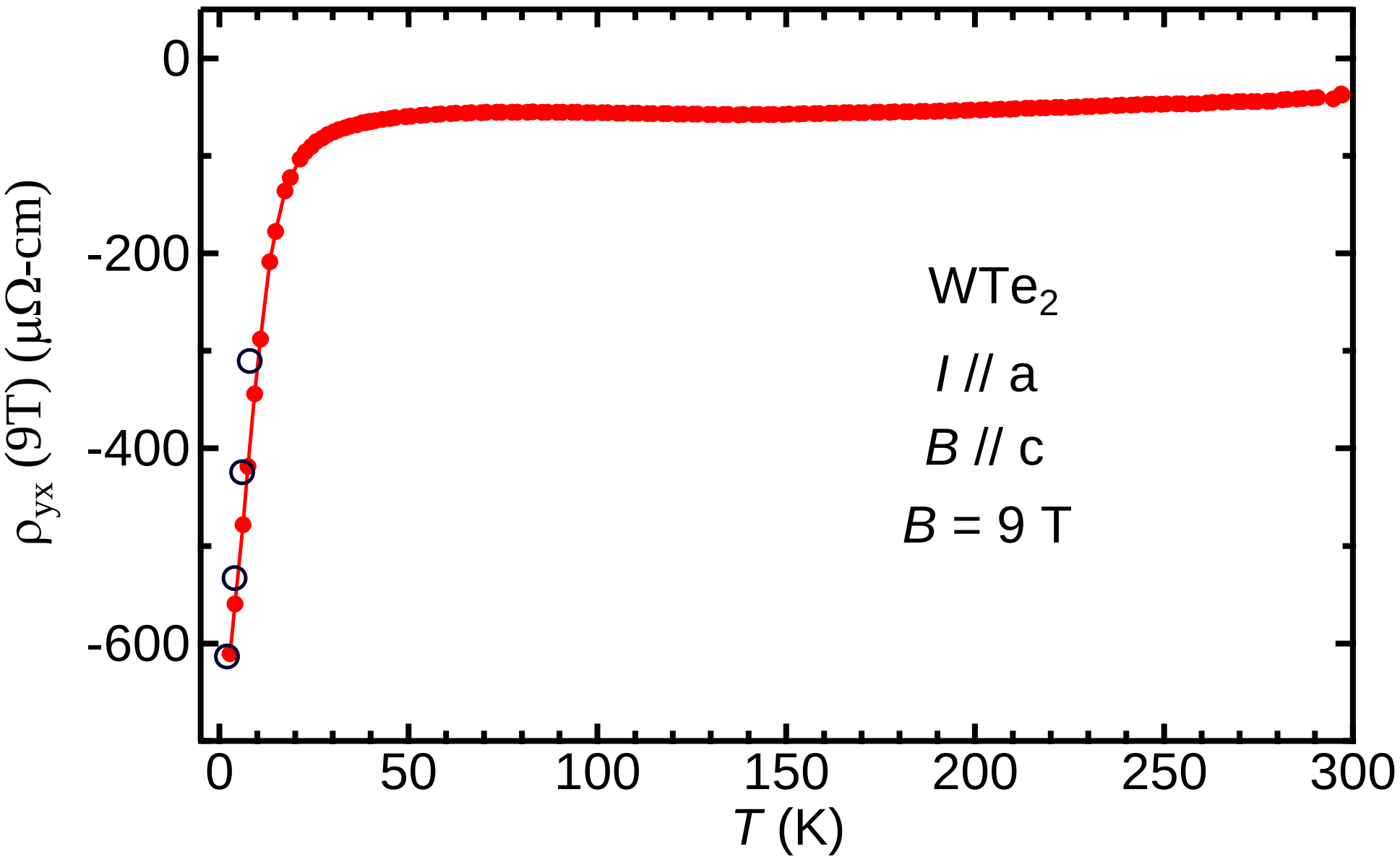

**Figure 2:** Temperature dependence of Hall resistivity ($\rho_{yx}$) for *I*//*a* and *B*//c in 9 T. The open circles represent two-band model fitting results.

Figure 3.

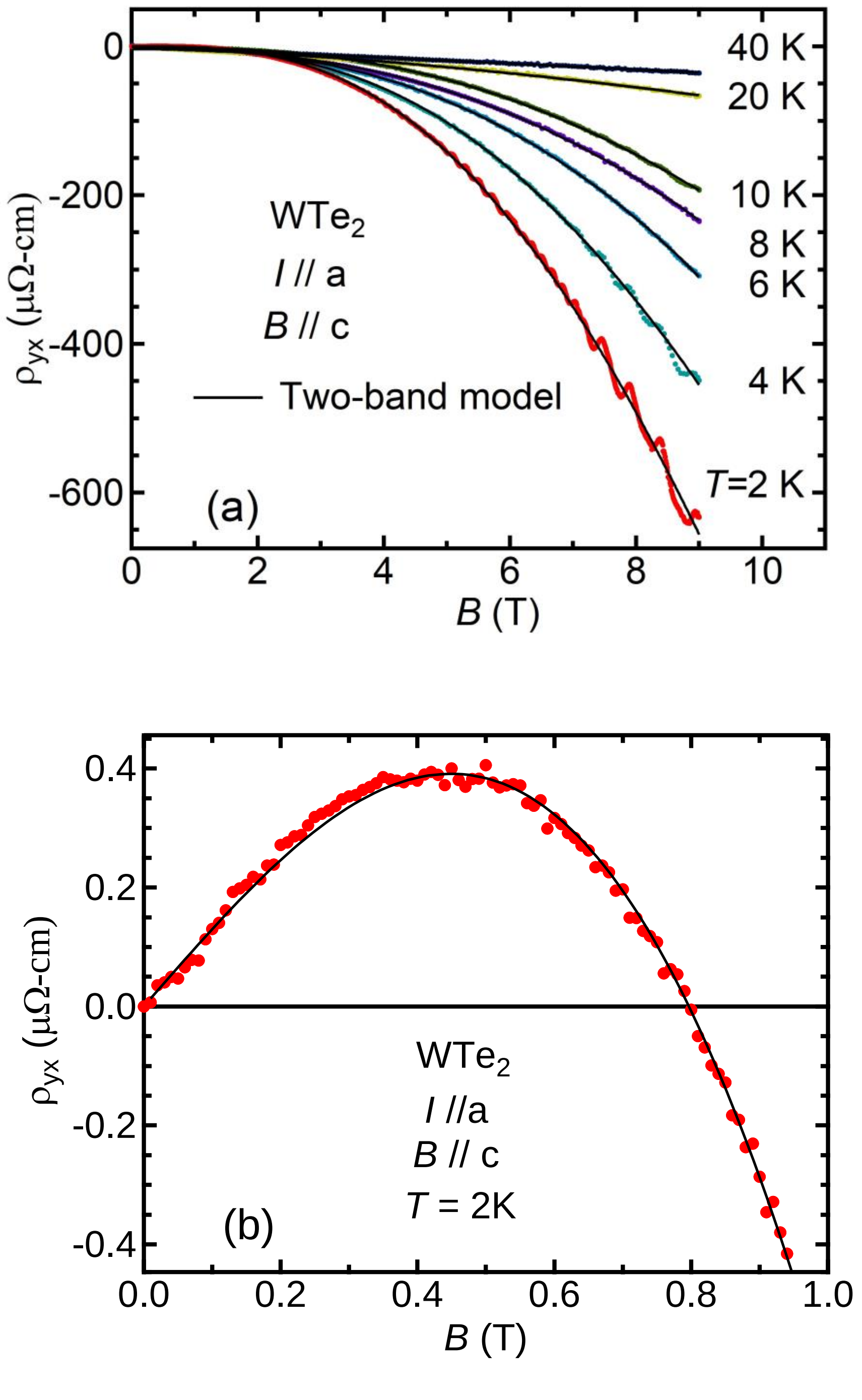

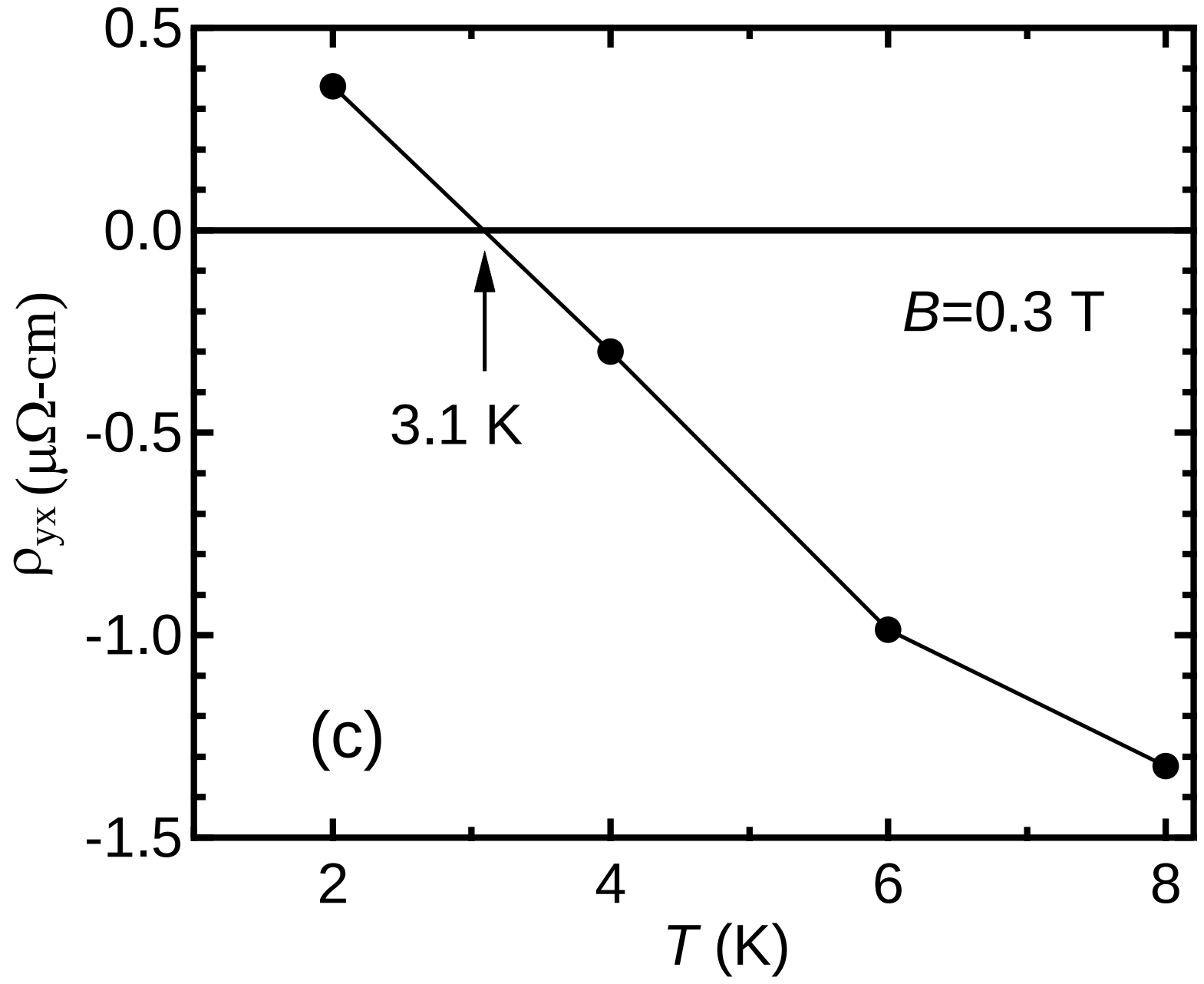

**Figure 3:** (a) Field dependence of the Hall resistivity at various temperatures for *I*//*a* and *B*//c. Two-band model fitting represented by solid lines shows a good agreement with the $\rho_{yx}$ data. (b) Expanded view of $\rho_{yx}$ vs *B* in a low field region at 2 K. Two-band model fitting represented by a solid line shows a good agreement with the $\rho_{yx}$ data. (c) $\rho_{yx}$ vs *T* in *B*=0.3 T. Data taken from Fig. 3(a).

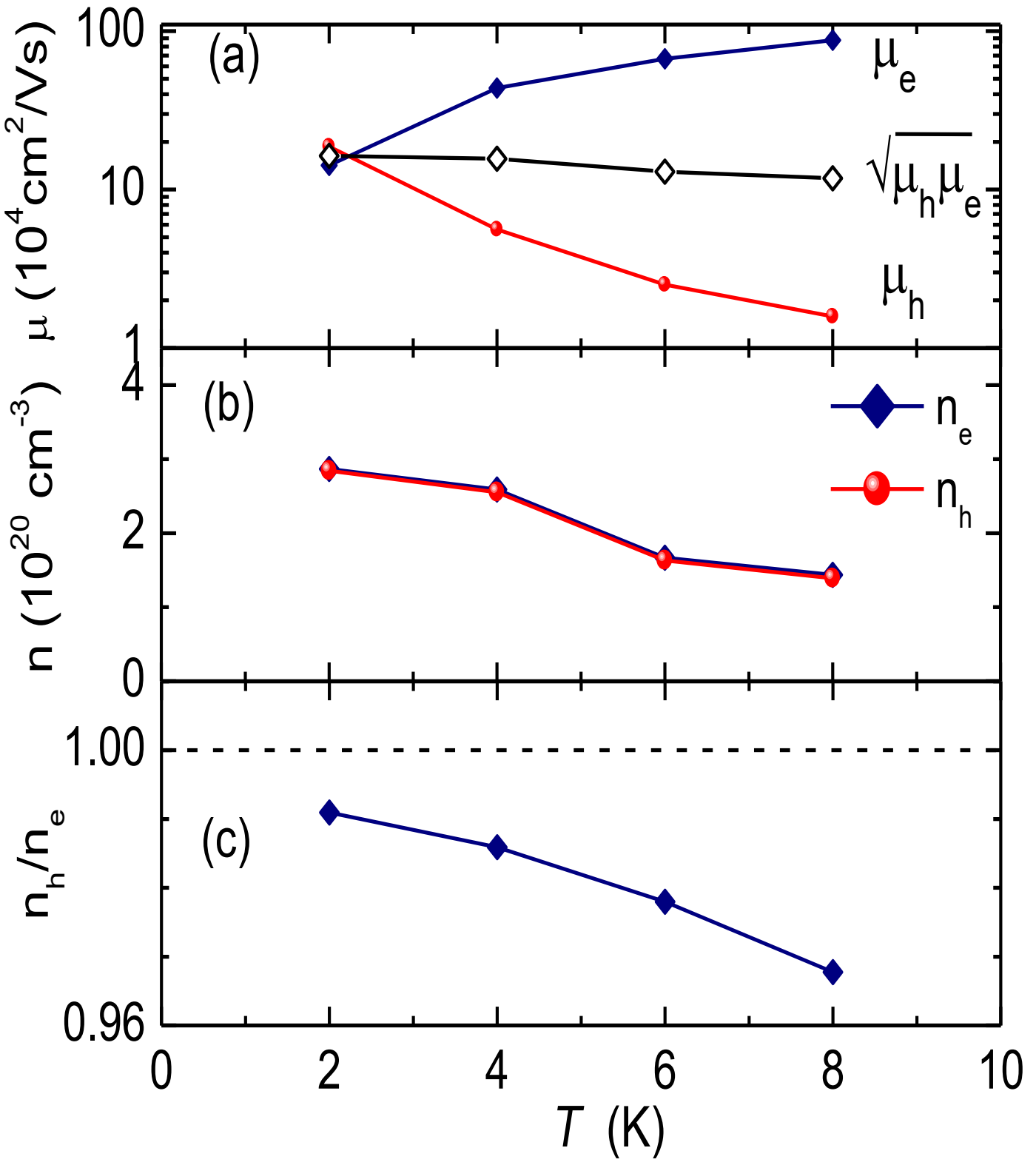

**Figure 4:** Temperature dependences of carrier mobility $\mu_e$, $\mu_h$, and the geometric-mean mobility $(\mu_e\mu_h)^{1/2}$ (a), carrier densities $n_e$ and $n_h$ (b), and the carrier density ratio $n_h/n_e$ (c). The obtained parameters are considered to represent those in the high-field region because the experimental data in 1 T<$B$<9 T are used.